\pdfoutput=1
\PassOptionsToPackage{expansion=false}{microtype}
\documentclass[sigconf,nonacm]{acmart}

\usepackage{array}
\usepackage{booktabs}
\usepackage{amsmath}
\usepackage{graphicx}
\usepackage{balance}
\graphicspath{{paper/figures/}{figures/}{./}}
\title{Which Eviction Policy Should an LLM Cache Use?
A Systematic Study Across Workloads,
Capacities, and Encoders}

\author{Yash Kulkarni}
\affiliation{
  \institution{University of Michigan}
  \city{Ann Arbor}
  \state{Michigan}
  \country{USA}
}
\email{yashkulk@umich.edu}

\author{Shubham Harkare}
\affiliation{
  \institution{University of Michigan}
  \city{Ann Arbor}
  \state{Michigan}
  \country{USA}
}
\email{sharkare@umich.edu}

\author{Arvind Yogesh Suresh Babu}
\affiliation{
  \institution{University of Michigan}
  \city{Ann Arbor}
  \state{Michigan}
  \country{USA}
}
\email{savyo@umich.edu}

\keywords{semantic caching, LLM serving, cache eviction,
comparative evaluation, approximate nearest neighbor, HNSW,
LMSYS, LLM-as-judge, quality-adjusted hit rate}

\setcopyright{none}
\acmConference[CSE 584]{CSE 584: Advanced Database Systems, Course Project}{Winter 2026}{Ann Arbor, MI}
\acmISBN{}
\acmDOI{}
\acmPrice{}
\begin{document}
\begin{abstract}
Semantic caches reuse an LLM response when the incoming query embedding lies
near a cached query, but proposed eviction policies have rarely been compared
under one protocol. Using CLEVER, we evaluate FIFO, LRU, LFU, ARC, GDSF, a
single-pass streaming adaptation of SISO, and a semantic-redundancy policy
across three ordered, deduplicated query corpora, three cache capacities, and
two encoders. No evaluated policy improves on LFU by more than 0.041
percentage points in any of the eighteen settings. Replacement is not
irrelevant: FIFO and streaming SISO trail LFU by as much as 8.67 and 8.55
points, respectively, at tight capacity.

We explain the missing upside with a conditional packing result. Under exact
lookup and insert-on-miss, a newly inserted entry cannot have a resident
neighbor within the hit radius, so a geometry-aware eviction rule receives
little new redundancy signal. A separate audit exposes a larger problem with
the evaluated operating point. At MiniLM's median nearest-neighbor threshold,
only 2.1--3.9\% of sampled LMSYS and QQP hits are judged
answer-substitutable, reducing raw hit rates of 51--60\% to
quality-adjusted rates of 1.1--2.2\%. The cross-encoder study further shows
that thresholds do not transfer between embedding models. LFU is the strongest
simple default in this protocol; deployment decisions should first establish
answer validity and then test sub-point policy differences with exact search.
\end{abstract}
\maketitle

\section{Introduction}
LLM inference incurs serving cost and latency \cite{kwon2023pagedattention}.
Semantic caching embeds an incoming query, searches previously answered
queries, and bypasses the model when a match falls within a distance threshold.

A semantic cache makes three relevant decisions:
\begin{enumerate}
\item \textbf{Indexing:} Efficiently storing and retrieving high-dimensional vectors.
\item \textbf{Routing (Thresholding):} Deciding the distance threshold at which a cached response is a valid substitute for a fresh LLM call.
\item \textbf{Eviction:} Deciding which entries to remove when the cache reaches capacity.
\end{enumerate}

Eviction has attracted the widest range of proposals. Systems such as
GPTCache \cite{bang2023gptcache} default to LRU or LFU, which treat keys as
exact matches. Semantic policies instead protect isolated embeddings and
remove entries with nearby substitutes. This family includes SISO
\cite{kim2025siso}, a bandit formulation \cite{liu2026semantic},
learning-augmented agent-memory replacement \cite{sun2026solar}, and
semantic-aware prefix eviction \cite{fang2026prefix}. ARC
\cite{megiddo2003arc} and GDSF \cite{cherkasova1998gdsf} combine recency and
frequency signals without using embedding geometry.

Each policy has been evaluated on a different combination of corpus, capacity,
embedding model, and baseline. Published orderings can point in opposite
directions. Sun et al.\ \cite{sun2026solar}, for example, report LRU and LFU
losing to FIFO on agent-memory workloads; in our matrix FIFO never beats LRU.
A shared protocol makes the disagreement visible, but our data does not isolate
its cause.

We built CLEVER so that every policy sees the same ordered corpus, index,
threshold, measured suffix, and cache capacity. We compare FIFO, LRU, LFU,
ARC, GDSF, streaming SISO, and an instrumented semantic-redundancy policy
across three corpora, three capacities, and two encoders. We also audit whether
the cached answer can be reused for the incoming query, because raw hit rate
does not distinguish useful reuse from a permissive threshold. ANN-index and
router benchmarks characterize the rest of the serving path.

No evaluated policy improves on LFU by more than 0.041\,pp, although FIFO
and streaming SISO lose several points at tight capacity. Insert-on-miss admission helps
explain why semantic policies fail to add value: under exact lookup, each new
miss enters outside every resident entry's hit radius. This packing condition
does not cover arbitrary prefill, approximate-search errors, or policies that
change stored representations. The quality audit then shows that eviction is
not the first deployment question. Thresholds change meaning across encoders,
and the MiniLM operating point used to stress eviction admits few
answer-substitutable LMSYS or QQP hits.

\textbf{Contributions.} This paper makes the following contributions:
\begin{enumerate}
\item \textbf{A Controlled Policy Comparison:} A head-to-head evaluation of FIFO, LRU, LFU, ARC, GDSF, streaming SISO, and a semantic redundancy policy under one protocol. No policy improves on LFU by more than 0.041\,pp, while FIFO and streaming SISO lose up to 8.67 and 8.55\,pp.
\item \textbf{A Conditional Mechanism:} A packing result for exact insert-on-miss caches, together with cache-density measurements from the 10\%-capacity MiniLM LRU configuration. It states when admission suppresses the redundancy signal and names the cases it does not cover.
\item \textbf{Per-Encoder Threshold Calibration:} Evidence that a MiniLM threshold yields a degenerate 100\% hit rate under gte-base. Quantile recalibration restores a usable experiment, but absolute LFU hit rates move by as much as 19\,pp.
\item \textbf{Quality-Adjusted Hit Rate:} A directional audit of answer substitutability. At the evaluated MiniLM threshold, quality-adjusted hit rates fall from 51--60\% to 1.1--2.2\% on LMSYS and QQP. Cross-runtime agreement within one judge family is $\kappa=0.826$ over 1{,}931 pairs.
\item \textbf{Deployment Guidance:} A decision procedure (Section~\ref{sec:guidance}) that separates the supported recommendation, LFU as the simple baseline in this protocol, from the measurements still needed for deployment.
\item \textbf{Serving-Path Benchmarks:} A comparison of FAISS index structures (Flat, HNSW, IVF, LSH) on 499K vectors establishing HNSW as Pareto-optimal (Recall@1 $= 0.989$ at $P50 = 0.52$ ms, a $34\times$ speedup over exact search for a 1.1\,pp recall loss), and a cost-based adaptive router achieving 60.33\% mean nominal latency savings across five seeds.
\end{enumerate}

\section{Related Work}
\subsection{Semantic Caching for LLMs}
GPTCache \cite{bang2023gptcache} provides an open-source architecture for
embedding queries and searching vector indexes such as FAISS
\cite{johnson2019faiss}, with static thresholds and LRU/LFU eviction. Later
systems move the cache client-side for privacy \cite{gill2025meancache},
cluster chat queries \cite{li2024scalm}, add conversation context
\cite{yan2025contextcache}, or learn per-entry thresholds with an error-rate
guarantee \cite{schroeder2025vcache}. Liu et al.\
\cite{liu2026semantic} formulate eviction as a combinatorial bandit. None of
these papers compares the candidate policies under one protocol.

\subsection{ANN Indexing and Vector Search}
Semantic caching requires similarity search over high-dimensional embeddings.
FAISS \cite{johnson2019faiss} implements several ANN indexes. HNSW
\cite{malkov2018hnsw} uses hierarchical graph navigation to trade recall for
latency. LSH \cite{indyk1998ann} uses random projections; our benchmark finds
lower recall under the tested $L_2$ thresholds. Product Quantization
\cite{jegou2011pq} compresses embeddings when memory is constrained.

\subsection{Cache Eviction Policies}
Belady's algorithm \cite{belady1966study} is the offline upper bound that
evicts the entry used furthest in the future. LRU-K \cite{oneil1993lruk}
tracks the $K$th most recent access to resist sequential scans. ARC
\cite{megiddo2003arc} balances recency and frequency with two adaptive LRU
lists. These policies operate on exact-match keys; our comparison asks whether
embedding-aware replacement improves on that foundation.

\subsection{Comparative Evaluations and Contemporaneous Work}
Several recent papers evaluate multiple replacement policies on semantic
workloads, and the picture they collectively paint is inconsistent, which is
part of what motivates a controlled comparison. Sun et al.\
\cite{sun2026solar} study eight policies on agent retrieval buffers and
report that LRU and LFU consistently underperform a naive FIFO baseline,
attributing this to absent temporal locality and frequency concentration in
their workloads. They manage stored agent experience, whereas we cache
query-to-response pairs from ordered, deduplicated corpora. Our ordering is the
opposite, with FIFO below LRU in every setting. Repetition, locality, cache
semantics, and implementation could each explain the difference; the present
experiments do not separate them. We include FIFO so the ordering can be
compared directly (Section~\ref{sec:matrix}).

Fang et al.\ \cite{fang2026prefix} learn semantic-aware eviction for prefix
caches, operating on KV blocks inside the serving engine rather than on
completed query-response pairs; their reported reuse variation across token
types is a different phenomenon from the query-level redundancy we measure.
Baral et al.\ \cite{baral2026calibration} address threshold calibration in
semantic caching and argue that ranking metrics such as PR-AUC misrepresent
behavior at a fixed deployment threshold, proposing precision-at-cache-hit
metrics instead. Their work and ours reach a similar conclusion, that a
permissive threshold inflates apparent cache performance, from opposite
directions. They evaluate nine bi-encoders independently on a static labeled
sentence-pair benchmark, determine hit validity from ground-truth pair
labels, and study neither eviction nor the transfer of a calibrated
threshold from one encoder to another. We run a streaming cache simulation
in which admission and eviction interact, judge validity with an LLM on
hits the deployed threshold actually admitted, and show that transferring a
threshold across encoders fails outright. The two approaches are
complementary: their metrics describe how to choose a threshold, our
measurements describe what happens downstream when one is chosen badly. To our knowledge no prior study places the
classical policies, the LLM-serving-specific policies, and a semantic policy
under one protocol across workloads, capacities, and encoders, or pairs
that comparison with an audit of whether the measured hits are valid.

\subsection{Retrieval-Augmented Generation and Non-Parametric Memory}
Retrieval-augmented generation conditions model output on external context
\cite{lewis2020rag}. kNN-LM \cite{khandelwal2020knnlm}, RETRO
\cite{borgeaud2022retro}, and Atlas \cite{izacard2023atlas} show benefits from
retrieval under different model and training settings; REPLUG
\cite{shi2023replug} and Self-RAG \cite{asai2024selfrag} add black-box and
adaptive variants.

Retrieval is not universally beneficial. Mallen et al.\ \cite{mallen2023trust}
show that retrieval augmentation can hurt performance on facts well memorized
by parametric models, and Liu et al.\ \cite{liu2024lost} show that LLMs often
fail to use retrieved context placed in the middle of long inputs. These
results motivate measuring cache behavior by workload rather than assuming
that every retrieved match helps.

\section{System Architecture and Methodology}
CLEVER is organized as a sequence of modular components: an index benchmark, a cost-based router, an eviction layer hosting seven interchangeable policies, and an LLM-based audit of hit quality. This section describes each in turn.

\subsection{Index Benchmarking}
We compare four FAISS \cite{johnson2019faiss} index structures:
\begin{itemize}
\item \textbf{Flat (Exact):} Exhaustive $L_2$ distance search, guaranteeing 1.0 recall.
\item \textbf{HNSW (Hierarchical Navigable Small World) \cite{malkov2018hnsw}:} A graph-based approach optimizing for rapid routing and high recall.
\item \textbf{IVF (Inverted File Index):} A partition-based approach optimizing for memory footprint and throughput via Voronoi cells.
\item \textbf{LSH (Locality-Sensitive Hashing):} Evaluated for extreme memory-constrained environments.
\end{itemize}

\subsection{Cost-Based Adaptive Router}
We evaluate a router that selects an $L_2^2$ distance threshold
($\theta$) from a calibration set by trading estimated serving cost against
embedding similarity.

The router sweeps candidate thresholds to maximize a combined objective function:
\begin{equation}
\mathit{Obj} = \mathit{CostSavings}(\theta) - \lambda(1 - \mathit{Quality}(\theta))
\end{equation}
where $\mathit{CostSavings}$ comes from the hardware cost model and
$\mathit{Quality}$ is a cosine-similarity proxy
($1 - \frac{d^2}{2}$, where $d$ is the $L_2$ distance
between normalized vectors). We set $\lambda=2.0$. Normalization makes the
distance-to-cosine conversion exact; it does not make cosine similarity an
exact measure of answer quality. Section~\ref{sec:qadj-meaning} audits this
limitation.

\subsection{The Semantic Redundancy Policy}
The semantic policy in the comparison, from which CLEVER (Cluster-Level Eviction for Vector Embedding Retrieval) takes its name, scores entries by the ratio of their semantic redundancy to their historical utility.

\textbf{The Scoring Function:}
\begin{equation}
\mathit{score}(e) = \frac{r(e) + \mu}{\alpha \cdot \mathit{recency}(e) + \beta \cdot \mathit{frequency}(e) + \varepsilon}
\end{equation}

\begin{itemize}
\item \textbf{Redundancy $r(e)$:} The fraction of active cached entries within distance $\theta_e$ of $e$.
\item \textbf{$\mu$-Smoothing:} An additive Laplace-style constant. As $\mu \to \infty$, the semantic term is washed out, and the ordering approaches pure inverse-utility (LFU/LRU). This guarantees the policy degrades gracefully rather than failing entirely if the redundancy signal is absent.
\item \textbf{Recency Floor:} $recency(e)$ is ranked and floored at $1/N$. This prevents the oldest entry from driving the denominator to $\varepsilon$, which would cause its score to explode and render it ``immortal'' regardless of its redundancy.
\end{itemize}

\textbf{Incremental Symmetric Redundancy Graph:}
Computing $r(e)$ naively requires $O(N^2)$ distance calculations per eviction. We implement an incremental approach:
\begin{itemize}
\item \textbf{On Insert:} We sample $S \le 1024$ active entries, compute distances to the new entry $e$, and \textit{symmetrically} add edges to the neighbor graph for any pairs within $\theta_e$. This bounds insertion to $O(S)$.
\item \textbf{On Evict:} We remove the evicted node from the neighbor lists of its connected peers in $O(|neighbors|)$ time.
\item \textbf{Batch Rebuilds:} To correct for sampling bias, we perform periodic full rebuilds of the graph via FAISS batch search.
\end{itemize}

\subsection{Classical and LLM-Serving Policies}
A comparison restricted to LRU, LFU, and the semantic policy would be too narrow to support deployment guidance, so we add four further policies spanning arrival-order, adaptive, cost-aware, and LLM-cache-specific designs. Each must be adapted to a semantic cache, where entries are unit-norm embeddings and cache ids are slot-unique (an evicted id is never re-requested).

\textbf{FIFO} evicts the oldest insertion and ignores accesses entirely. It is the weakest policy that still functions as a cache, and it anchors the bottom of the comparison: any policy that fails to beat it is extracting no value from the recency, frequency, or geometric signals it consults. We include it also because Sun et al.\ \cite{sun2026solar} report it outperforming LRU and LFU on semantic workloads, a claim that cannot be checked against our matrix without running it.

\textbf{ARC} \cite{megiddo2003arc} balances a recency list $T_1$ against a frequency list $T_2$ using ghost lists $B_1$/$B_2$ of recently evicted keys; a request for a ghosted key adapts the target split $p$. Because our cache ids are slot-unique, a classic exact-match ghost hit can never fire. Our adaptation detects ghost hits semantically: each ghost retains the evicted entry's embedding, and an inserted query whose embedding lies within
$L_2^2 \le 0.20$ of a ghost (cosine ${\approx}\,0.90$ for unit-norm MiniLM vectors) counts as a ghost hit. This is the semantic-cache analogue of re-requesting a recently evicted item. On a match, $p$ adapts with the classic deltas, the ghost is retired, and the new entry is admitted directly to $T_2$; ghost lists are LRU-bounded at capacity $c$.

\textbf{GDSF} \cite{cherkasova1998gdsf} assigns each entry the priority $L + \mathit{frequency}(e) \cdot \mathit{cost}(e)/\mathit{size}(e)$, where the inflation clock $L$ rises to each victim's priority (aging). We instantiate $\mathit{cost}(e) = 1/\lVert e \rVert$ and $\mathit{size}(e)$ as the embedding dimension. On unit-norm, fixed-dimension embeddings every entry has $\mathit{cost} \approx 1$ and identical size, so the cost/size term is uniform and GDSF reduces to \emph{LFU-with-aging}; the general formula is implemented regardless.

\textbf{Streaming SISO.} SISO \cite{kim2025siso} is a caching policy designed specifically for LLM serving. Each cached entry acts as a cluster centroid carrying a \emph{cluster size} (semantic locality, long-term value) and an \emph{access count} (short-term popularity). On insert, the closest cached entry within the clustering threshold
$\theta_C = 0.86$ cosine ($L_2^2 < 0.28$) absorbs one unit of cluster mass; an insert that opens a new semantic region receives temporary infinite-access-count protection. Every 1{,}000 requests a maintenance round decays all cluster sizes by $1.1\times$ and resets access counts; eviction removes the entry with the lexicographically smallest (cluster size, access count) pair. SISO's original cache manager re-clusters query logs offline; our implementation maps its merge step to online per-insert merging with the search bounded at 1{,}024 sampled entries, and drives its M/D/1-based adaptive retrieval threshold $\theta_R$ from a sliding-window hit rate. As in the original design, $\theta_R$ does not influence victim selection. We therefore evaluate this adapted, online single-pass variant, denoted \emph{streaming SISO} throughout, rather than the original offline re-clustering manager; its results should be read as characterizing the streaming adaptation, and a comparison against the authors' reference implementation is left to future work.

All seven policies (FIFO, LRU, LFU, Semantic, ARC, GDSF, streaming SISO) run under an identical protocol: the policy only observes insert, access, and evict events and selects victims, while admission, lookup, the
$L_2^2 = 0.90$ hit threshold, capacity-sized prefill, and final-70\%
measurement suffix are fixed by the cache. The judged hit-quality audit
(Section~\ref{sec:judge}) predates the FIFO runs and therefore covers the
other six policies only.

\subsection{Per-Encoder Threshold Calibration}
\label{sec:calibration}
Similarity thresholds are not portable across embedding models. Our MiniLM
configuration uses an $L_2^2$ hit threshold of $0.90$ (cosine $\geq 0.55$),
which sits at MiniLM's median nearest-neighbor cosine similarity ($0.553$)
over the 100K-query corpora. The second encoder, \texttt{gte-base} (768
dimensions) \cite{li2023gte}, is strongly anisotropic
\cite{ethayarajh2019contextual}: random query pairs
average cosine similarity $0.69$--$0.76$ across our three datasets, versus
approximately $0$ for MiniLM. A threshold of cosine $0.55$ therefore lies
below the random-pair mean under gte-base, and essentially every pair
of queries qualifies as a cache hit. The same numeric threshold therefore
does not define the same operating point across these encoders.

We therefore recalibrate every distance-bearing parameter for gte-base by
anchoring it at the same quantile of the gte similarity distribution that the
MiniLM value occupies in MiniLM's. The hit threshold becomes $0.30$ $L_2^2$
(cosine $0.85$), the gte nearest-neighbor median; the semantic policy's
redundancy threshold moves with it ($0.30$, preserving the invariant that two
cached entries are neighbors if and only if one would be a hit for the other).
The streaming-SISO
clustering and retrieval thresholds move from cosine $0.86$ to $0.94$
(the deep random-pair tail under both encoders), with its dynamic-threshold
floor raised from $0.60$ to $0.85$ so it cannot wander below gte's random-pair
mean ($\approx 0.74$). ARC's ghost-list match threshold tightens from $0.20$
to $0.10$ $L_2^2$ (cosine $0.95$): at $0.20$, ghost hits fired on
$47$--$89\%$ of merely topical gte neighbors, corrupting ARC's adaptation
signal. All non-geometric hyperparameters are unchanged.
This calibration uses the evaluated 100K corpora rather than a held-out set.
It makes the comparison non-degenerate; it does not establish that either
threshold admits valid answer reuse.

\subsection{Judged Hit Quality}
\label{sec:judge}
Hit rate treats every match under the threshold as a success, yet a
cached response is only useful if it actually answers the new query.
To measure this, we instrument the eviction driver to log every
served cache hit: the incoming query, the matched cached query, and
their $L_2^2$ distance, alongside the full run configuration. We
enable logging for the 10\%-capacity MiniLM runs (seed 42, 100k-query
streams) across all three datasets and six policies. From each
(dataset, policy) cell a stratified sampler draws 1{,}000 hits spread
across five quantile bins of $L_2^2$ distance, so borderline hits near
the threshold are represented rather than only easy near-duplicates;
this yields 18{,}000 records, or 13{,}669 unique query pairs after
deduplicating pairs sampled under multiple policies.

Each pair is scored by an LLM judge \cite{zheng2023judging}
under a fixed binary rubric. The system prompt casts the judge as a ``strict evaluator of
semantic cache hits'' that must ``judge meaning, not wording'' and
answer NO ``if reusing the original answer could mislead or omit
something the new question specifically asks for.'' The user prompt
presents both queries and asks: \emph{``Would a correct, complete
answer to the ORIGINAL question also be a correct and complete answer
to the NEW question? Reply with exactly one word: YES or NO.''} We
call a YES verdict \emph{answer-substitutable}. The test is directional:
it does not establish symmetric query equivalence. QQP provides no stored
responses, so a response-grounded audit is outside this experiment. The judge
does not see embedding distance.

The judge is Llama-3.1-8B-Instruct \cite{grattafiori2024llama} served
locally via Ollama's OpenAI-compatible endpoint at temperature 0 with
a 16-token completion cap; verdicts are parsed by exact YES/NO prefix,
and 13{,}667 of 13{,}669 pairs parse cleanly. We also send 1{,}931 pairs
to \texttt{llama-3.1-8b-instant} on Groq under the same rubric and decoding
configuration. Because both runtimes serve the same model family, their
agreement measures runtime consistency rather than judge correctness.

\subsection{Hyperparameter Configuration}
Table~\ref{tab:hyperparams} summarizes the semantic-policy and router
parameters. Within each reported matrix, values are fixed across datasets,
capacities, and seeds. Distance parameters change by encoder as described in
Section~\ref{sec:calibration}; that in-sample calibration is not a held-out
hyperparameter study. The smoothing constant $\mu$ and recency floor keep the
policy defined when redundancy is weak (Section~\ref{sec:mu}).

\begin{table}[h]
\caption{Hyperparameter configuration for the semantic
eviction policy and the adaptive router.}
\label{tab:hyperparams}
\small
\begin{tabular}{@{}lllp{2.6cm}@{}}
\toprule
Parameter & Symbol & Value & Justification \\
\midrule
Recency weight      & $\alpha$      & 1.0        & Recency in utility denom. \\
Frequency weight    & $\beta$       & 1.0        & Frequency in utility denom. \\
Denominator floor   & $\varepsilon$ & $10^{-9}$  & Prevents division by zero \\
Smoothing constant  & $\mu$         & 0.1        & Laplace-style smoothing \\
Neighbor sample cap & $S$           & 1024       & Bounds $O(S)$ insert cost \\
Penalty coefficient & $\lambda$     & 2.0        & Suppresses false accepts \\
Routing threshold   & $\theta_r$    & 0.772      & Adaptive mean, 5 seeds \\
Eviction threshold  & $\theta_e$    & 0.90       & Semantic neighbor cutoff \\
\bottomrule
\end{tabular}
\end{table}

\section{Experimental Setup}
All experiments share the following data, encoders, and measurement protocol.
\begin{itemize}
\item \textbf{Datasets:} We extract the first user utterance from each LMSYS-Chat-1M conversation \cite{zheng2023lmsys}, individual questions from Quora Question Pairs \cite{iyer2017qqp}, and instruction prompts from MOSS \cite{sun2024moss}. Preprocessing removes exact duplicate text. The original three-policy evaluation uses the resulting 579{,}753-query LMSYS corpus. For the policy matrix, we sample 100{,}000 rows from each processed corpus without replacement and restore source-row order.
\item \textbf{Embeddings:} Queries are encoded into 384-dimensional vectors using the \texttt{all-MiniLM-L6-v2} sentence-transformer model \cite{reimers2019sbert}. The cross-encoder replication re-embeds all three datasets with \texttt{gte-base} (768 dimensions) \cite{li2023gte}.
\item \textbf{Normalization:} All vectors are $L_2$-normalized, so that squared $L_2$ distance and cosine similarity are interchangeable via $L_2^2 = 2(1-\cos) = 2 - 2\cos$ (equivalently $\cos = 1 - L_2^2/2$), as required by the adaptive router.
\item \textbf{Workload Order:} Queries follow processed-corpus row order. These inputs are ordered corpora, not production request traces. Exact query repetition is absent because the pipeline keeps only the first occurrence of each query text; recorded occurrence counts are not replayed.
\item \textbf{Multi-Turn Handling:} LMSYS conversation context and later turns are excluded from the cache key. The evaluation therefore measures first-turn, single-query matching.
\item \textbf{Cache Capacity Sweep:} We evaluate cache sizes at 10\%, 20\%, and 30\% of the total unique query volume.
\item \textbf{Lookup Index:} The eviction matrix uses FAISS HNSW with $M{=}32$, $\mathit{efConstruction}{=}128$, and $\mathit{efSearch}{=}128$ for every policy.
\item \textbf{Hardware:} All experiments run on the University of Michigan Great Lakes HPC cluster. The resources we report are the SLURM allocations requested per job, not the hardware of the hosting node: index benchmarking requests 8 CPU cores and 64\,GB RAM on the Standard partition, the five-seed adaptive-routing evaluation 8 cores and 32\,GB on the GPU partition, and the computationally intensive eviction multi-seed matrix 16 cores and 64\,GB on the Large-Memory partition. The requested allocations are recorded in the accompanying \texttt{slurm/} batch scripts; each run's manifest separately records the physical hardware of the larger shared node that hosted the job.
\item \textbf{Prefill and Measurement:} For a capacity fraction $C$, the cache is built from the first $C$ fraction of the corpus. The remainder of the first 30\% is excluded, and hit rates cover the final 70\%. This keeps the measured suffix fixed across capacities, but it is a capacity-sized prefill rather than a 30\% online warmup.
\item \textbf{Repetitions:} Eviction cells use seeds 42, 123, and 456. These seeds vary randomized policy-side sampling and check repeatability; they do not provide independent traces. Routing uses five seeds, and the judge sample comes from the seed-42 runs.
\end{itemize}

\section{Results}
We report the serving-path benchmarks first, then the policy comparison, its cross-encoder replication, and the judged audit of hit quality.

\subsection{Index Benchmarking}
\begin{figure}[h!]
  \centering
  \includegraphics[width=\columnwidth]{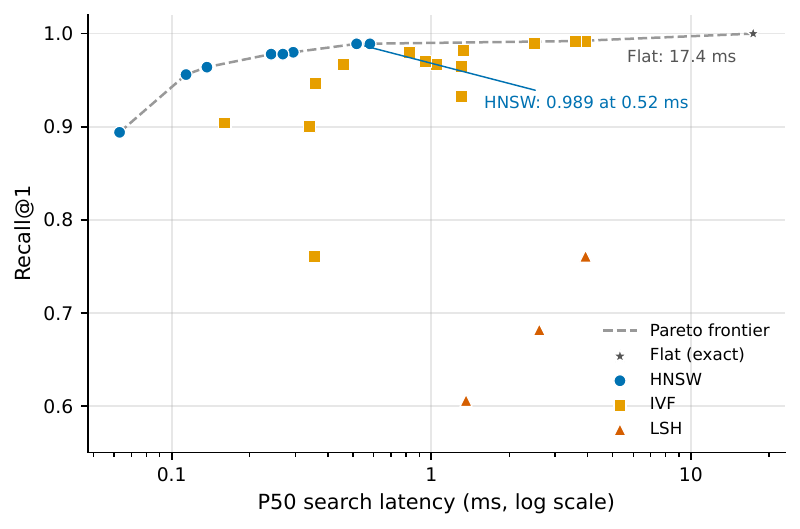}
  \caption{Recall@1 vs.\ P50 search latency (log scale) for all
  index configurations on 499K vectors, uniform workload. HNSW
  traces the Pareto frontier up to Recall@1 $=$ 0.989 at 0.52 ms;
  exact Flat search pays 17.4 ms for perfect recall.}
  \label{fig:hnsw}
\end{figure}

Index benchmarking was conducted on 499K vectors drawn from the full embedding corpus, under a uniform query workload. At this scale, HNSW ($M{=}32$, $\mathit{efConstruction}{=}256$, $\mathit{efSearch}{=}256$) achieves a Recall@1 of 0.989 with a P50 search latency of 0.52 ms, versus 17.4 ms for exact Flat search: a $34\times$ speedup for a 1.1\,pp recall loss. IVF reaches comparable recall only at high probe counts (0.989 at $\mathit{nlist}{=}1024$, $\mathit{nprobe}{=}128$), where its P50 rises to 2.5 ms, roughly five times HNSW's, and it degrades sharply at low probe counts (0.761 at $\mathit{nprobe}{=}1$). LSH does not exceed 0.761 recall even at 1,536 bits, rendering it non-viable for semantic caching. HNSW traces the Pareto frontier up to Recall@1 $= 0.989$; beyond that point, only near-exhaustive scans buy the last fraction of recall, at an order-of-magnitude latency cost (Fig.~\ref{fig:hnsw}).

\subsection{Adaptive Routing}
\begin{figure}[t]
  \centering
  \includegraphics[width=\columnwidth]{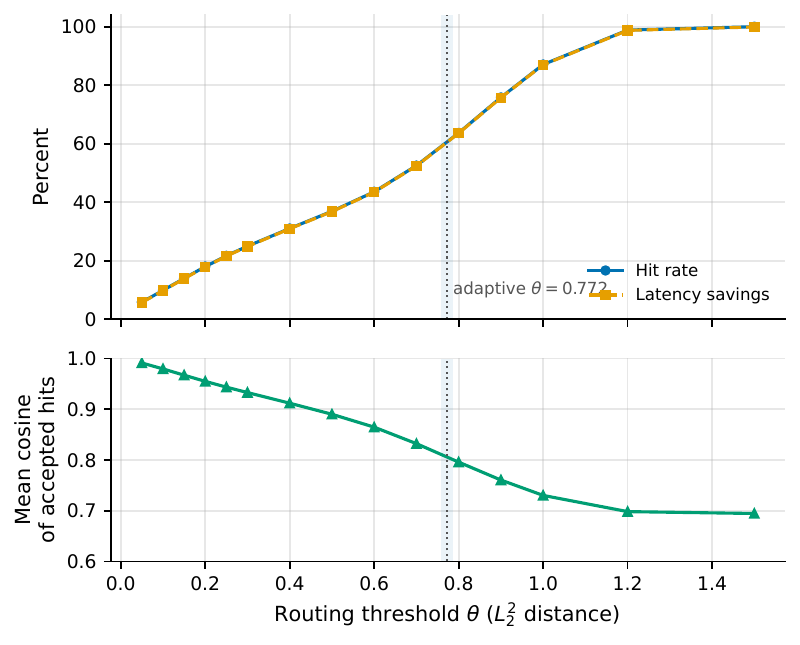}
  \caption{Five-seed mean threshold sweep under random cache
  fill. Hit rate and latency savings rise as the threshold
  loosens while accepted-hit quality falls; the adaptive router
  settles at $\theta = 0.772 \pm 0.015$ (shaded band).}
  \label{fig:router}
\end{figure}

The router selects $\theta\in[0.76,0.79]$. Under random cache fill, the
five-seed mean is $\theta=0.772\pm0.015$, with a 60.43\% hit rate, mean
accepted-hit cosine of 0.806, and 60.33\% nominal latency savings. In the
sweep (Fig.~\ref{fig:router}), hit rate and savings rise as the threshold
loosens while accepted-hit cosine falls. The mean cosine drops from 0.76 at
$\theta=0.9$ to 0.69 at $\theta=1.5$, where every query is accepted.

\subsection{Cache Eviction Performance}
\begin{figure}[t]
  \centering
  \includegraphics[width=0.88\columnwidth]{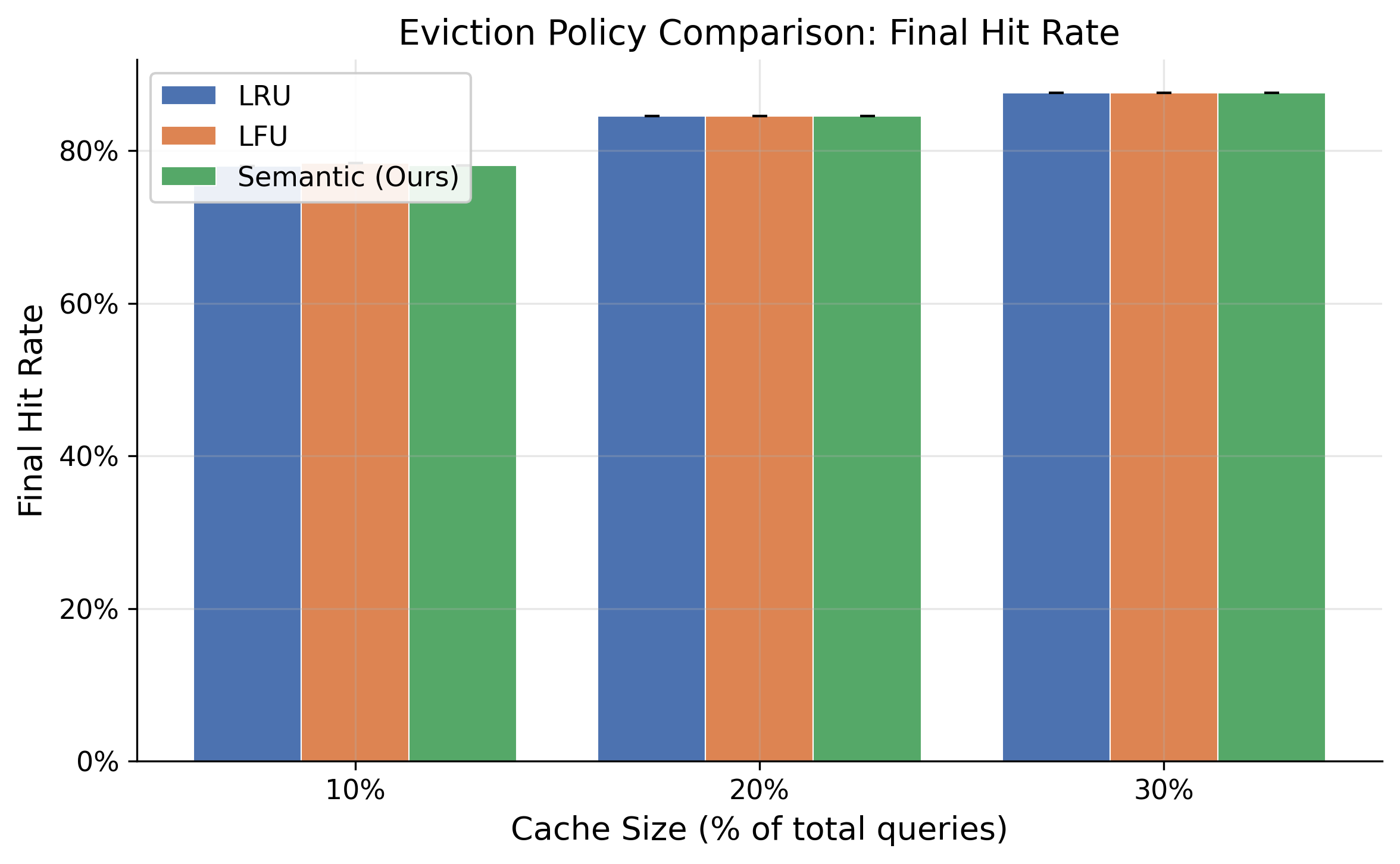}
  \caption{Hit rate across eviction policies at three cache
  capacities on the full processed LMSYS corpus.}
  \label{fig:eviction}
\end{figure}

The semantic eviction policy provides no hit-rate gain on the full LMSYS corpus
(Fig.~\ref{fig:eviction}). At 10\% capacity, LFU leads the semantic policy by
0.26\,pp and LRU by 0.33\,pp; at 20\% and 30\%, all three values agree to
four decimal places. The seeds exercise internal randomized sampling rather
than new corpora, so Table~\ref{tab:hit-rate} reports repeatability, not
sampling uncertainty over request traces.

\begin{table}[t]
\caption{Cache hit rate (mean $\pm$ std) across three seeds
(42, 123, 456) at three cache capacities. Differences across
policies are at most 0.34\,pp in this experiment.}
\label{tab:hit-rate}
\resizebox{\columnwidth}{!}{%
\begin{tabular}{lccc}
\toprule
Cache Size & LRU & LFU & Semantic \\
\midrule
10\% & $0.7802 \pm 0.0001$ & $\mathbf{0.7835 \pm 0.0000}$
     & $0.7808 \pm 0.0000$ \\
20\% & $0.8453 \pm 0.0000$ & $0.8453 \pm 0.0000$ & $0.8453 \pm 0.0000$ \\
30\% & $0.8756 \pm 0.0000$ & $0.8756 \pm 0.0000$
     & $0.8756 \pm 0.0000$ \\
\bottomrule
\end{tabular}%
}
{\footnotesize $\dagger$ All stds $<$ 0.00005 except LRU at
10\% (std\,$=$\,0.00005); all round to $\pm$0.0000 or
$\pm$0.0001 as shown. Variance is non-zero but below
reporting precision.}
\end{table}

\paragraph{Runtime overhead.}
Table~\ref{tab:latency} reports mean per-query processing time for the same runs.

\begin{table}[t]
\caption{Mean per-query processing time across three seeds.
Semantic/LRU reports multiplicative overhead relative to LRU.}
\label{tab:latency}
\small
\begin{tabular}{@{}lcccc@{}}
\toprule
Cache Size & LRU & LFU & Semantic & Sem./LRU \\
\midrule
10\% & 1.45 ms & 2.53 ms & 8.47 ms & $5.83\times$ \\
30\% & 1.90 ms & 3.67 ms & 15.70 ms & $8.24\times$ \\
\bottomrule
\end{tabular}
\end{table}

In the same runs, mean semantic-policy eviction time is 25.75 ms at 10\%
capacity and 90.27 ms at 30\%. The added computation yields no hit-rate gain
over LFU in these runs.

\subsection{Seven-Policy Comparison Across Workloads and Capacities}
\label{sec:matrix}
\begin{figure}[t]
  \centering
  \includegraphics[width=\columnwidth]{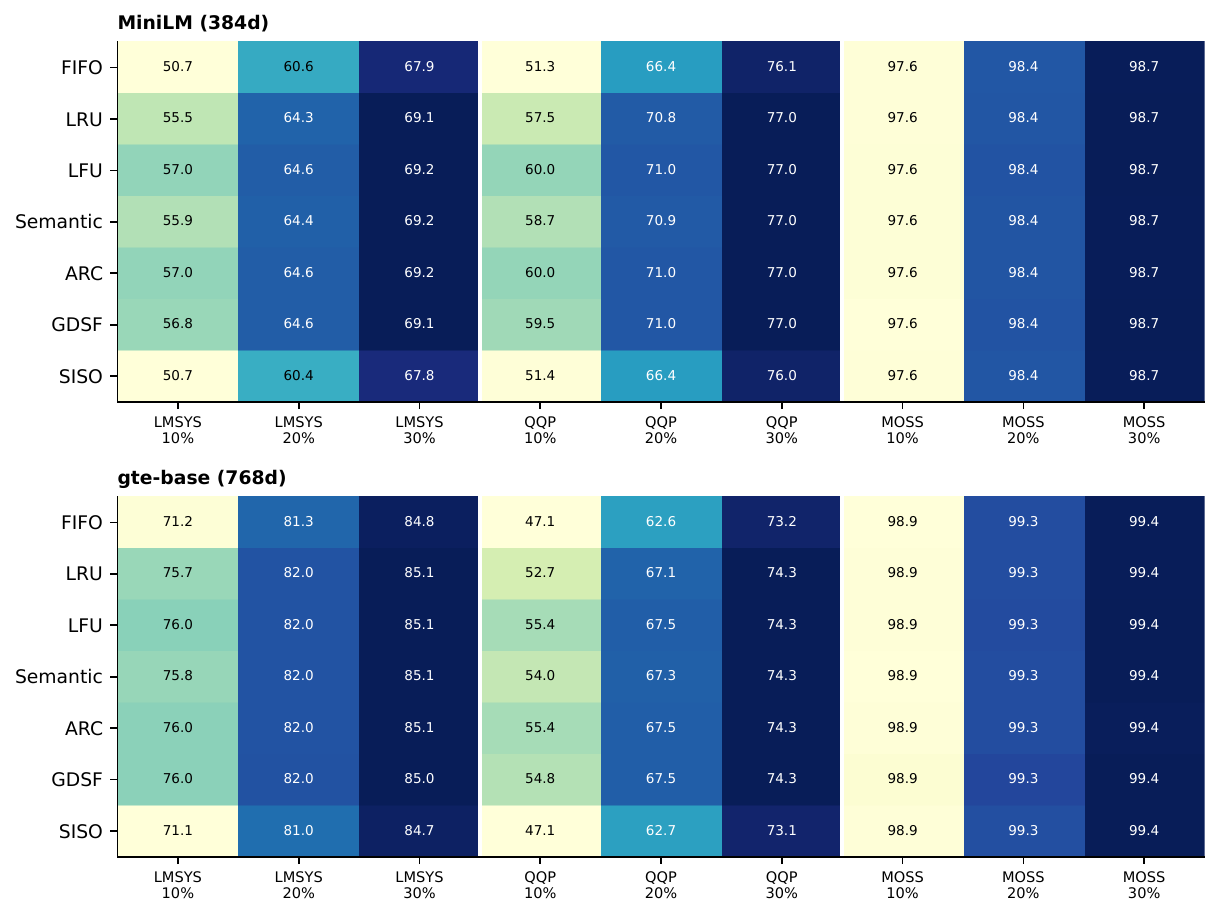}
  \caption{Hit rate (\%) for all seven policies across three
  datasets and three cache capacities, MiniLM (top) and
  gte-base (bottom) embeddings; color is normalized within
  each dataset block. No policy improves on LFU by more than
  0.041\,pp. FIFO and streaming SISO lose several points at tight
  capacity.}
  \label{fig:heatmap}
\end{figure}

\begin{figure}[t]
  \centering
  \includegraphics[width=\columnwidth]{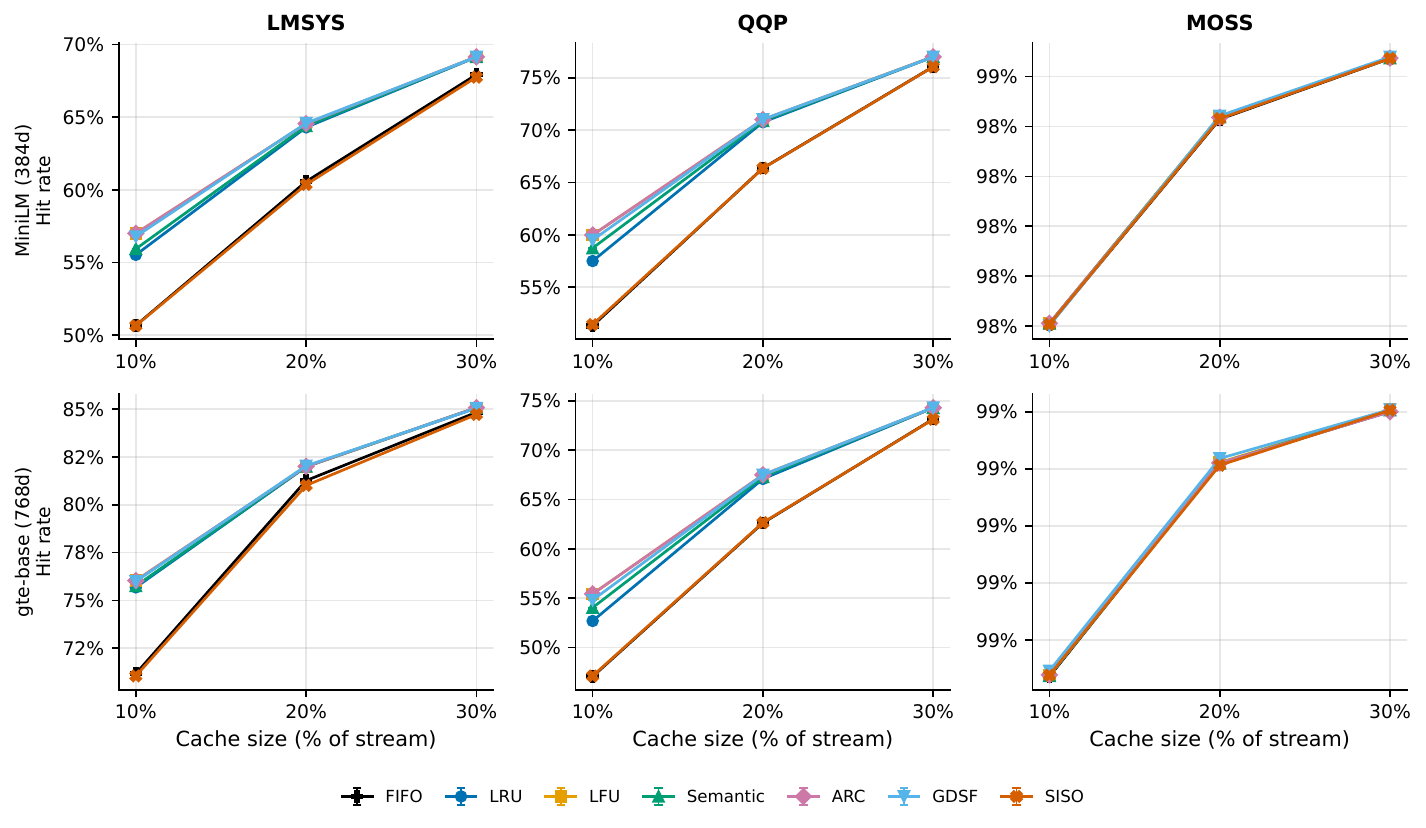}
  \caption{Cache-size ablation. Gaps narrow as capacity grows,
  especially among the leading policies, under both encoders.}
  \label{fig:ablation}
\end{figure}

We extend the eviction evaluation to the seven policies on the three ordered,
deduplicated 100{,}000-query corpora described above. QQP supplies related
questions, while templated MOSS instructions saturate the cache and act as a
control. Figure~\ref{fig:heatmap} reports the full matrix, and
Figure~\ref{fig:ablation} shows the capacity sweep.

Across all eighteen dataset, capacity, and encoder settings, the largest gain
over LFU is 0.041\,pp (GDSF on LMSYS at 20\% capacity under MiniLM), a
handful of requests in the measured suffix. At 10\% capacity, LFU reaches
57.0\% on LMSYS and 60.0\% on QQP. ARC remains within 0.01\,pp of LFU in
every setting. GDSF, which reduces to LFU-with-aging for fixed-size unit-norm
embeddings, trails LFU by at most 0.62\,pp (QQP at 10\% capacity under
gte-base) and leads it by at most 0.041\,pp.

FIFO trails LFU by 6.33, 3.97, and 1.22\,pp on LMSYS and by 8.67,
4.64, and 0.94\,pp on QQP at 10\%, 20\%, and 30\% capacity under
MiniLM, with the same pattern under gte-base. It does not beat LRU in any cell
and ranks sixth or seventh in sixteen; the two exceptions are saturated MOSS
cells under gte-base where all seven policies lie within 0.012\,pp, and in one
of them FIFO exactly ties LRU on total hits. This ordering differs from Sun et
al.'s agent-memory result \cite{sun2026solar}. Repetition, locality, cache
semantics, and implementation are plausible explanations, but this matrix
does not identify which property reverses the ordering.

The streaming SISO adaptation is the weakest performer on both sparse
corpora at every capacity, trailing LFU by 6.34, 4.18, and 1.40\,pp on
LMSYS and by 8.55, 4.66, and 0.96\,pp on QQP. Across those twelve
cells, streaming SISO and FIFO differ by 0.096\,pp on average and at most
0.254\,pp; streaming SISO is numerically lower in nine. Its cluster-size
signal is weak in the measured caches, while its maintenance rounds reset
access counts. The implementation also costs $2.8$--$5.2\times$ LRU's
per-query time across these cells.

Both deficits shrink with capacity, and the leading policies converge more
closely. LFU's lead over LRU falls from 1.5\,pp on LMSYS and 2.5\,pp on
QQP at 10\% capacity to at most 0.012\,pp at 30\%. The largest
streaming-SISO deficit at 30\% remains 1.40\,pp, so capacity narrows but does
not eliminate every policy gap.

MOSS behaves as a saturation control: all policies land between 97.6\%
(10\%) and 98.7\% (30\%), within 0.02\,pp at each capacity. The maximum
hit-rate standard deviation across the 126 cells is $3.1\times10^{-4}$ in
hit-rate units and reflects internal randomized sampling, not variation across
traces. At 10\% capacity, streaming SISO's mean processing time is
$3.5\times$ LRU's on LMSYS and $3.2\times$ on QQP; the semantic policy
costs $4.4\times$ and $3.8\times$.

\subsection{Cross-Encoder Replication}
\paragraph{Hit thresholds do not transfer across encoders.}
Running the gte-base matrix under the MiniLM-calibrated threshold ($0.90$
$L_2^2$) is degenerate: on all three datasets, all three seeds, and all six
policies evaluated at the time (the later FIFO runs use only the
recalibrated configuration), the final hit rate is exactly $1.0$ with zero
misses and zero evictions. Because gte-base's random-pair similarity exceeds the threshold,
every query ``hits'' whatever the cache contains, eviction never fires, and
the experiment measures nothing. We report this as a methodological result:
a semantic-cache deployment that swaps encoders without recalibrating its
threshold will accept the nearest resident entry for every request while
reporting a perfect hit rate.

\paragraph{Recalibrated replication.}
After per-encoder quantile recalibration, the leading-policy pattern returns. At
$10\%$ capacity the policy ordering on LMSYS and QQP is identical to MiniLM's:
$\text{ARC}=\text{LFU} > \text{GDSF} > \text{Semantic} > \text{LRU}$, with
streaming SISO and FIFO at the bottom. ARC and LFU match within $0.01$pp in
all nine dataset--capacity cells. The semantic policy again never beats LFU by more than $0.01$pp.
Streaming SISO trails the best policy by
$5.0$/$1.0$/$0.4$pp on LMSYS and $8.3$/$4.8$/$1.2$pp on QQP at $10/20/30\%$
capacity, with FIFO tracking streaming SISO to within $0.3$\,pp throughout; both
deficits narrow as capacity grows. The LFU--LRU frequency advantage likewise shrinks with
capacity ($+2.7$/$+0.4$/$+0.0$pp on QQP; $+0.4$/$+0.0$/$+0.0$pp on LMSYS),
and MOSS remains saturated for every policy ($0.989$--$0.994$ hit rate, fewer
than $750$ mean evictions). Figure~\ref{fig:ablation} shows both encoder rows.

\paragraph{Orderings transfer; absolute rates do not.}
While the leading-policy pattern recurs under both encoders, absolute hit rates do not:
at $10\%$ capacity, LFU achieves $0.760$ on LMSYS under gte-base versus
$0.570$ under MiniLM ($+19.0$pp), yet $0.554$ versus $0.600$ on QQP
($-4.6$pp). The differences are large and not even direction-consistent,
so hit rates measured under one encoder cannot be used to provision a cache
served by another. What does transfer is the relative conclusion:
none of the evaluated complex policies improves materially on LFU.

\subsection{Quality-Adjusted Hit Rate}
\begin{table*}[t]
\caption{Judged hit quality at 10\% cache (MiniLM, seed 42): raw hit
rate (\%), answer-substitutable (YES) fraction with Wilson 95\% CI
($n{=}1000$ per cell), and quality-adjusted hit rate
QA $=$ Raw $\times$ YES (\%). Within every dataset the YES intervals
of all six policies overlap.}
\label{tab:judge}
\centering
\small
\begin{tabular}{l rrr rrr rrr}
\toprule
 & \multicolumn{3}{c}{LMSYS} & \multicolumn{3}{c}{QQP} & \multicolumn{3}{c}{MOSS} \\
Policy & Raw & YES [95\% CI] & QA & Raw & YES [95\% CI] & QA & Raw & YES [95\% CI] & QA \\
\midrule
LRU & 55.5 & 3.6 [2.6, 4.9] & 2.0 & 57.5 & 2.5 [1.7, 3.7] & 1.4 & 97.6 & 25.1 [22.5, 27.9] & 24.5 \\
LFU & 57.0 & 3.9 [2.9, 5.3] & 2.2 & 60.0 & 2.6 [1.8, 3.8] & 1.6 & 97.6 & 26.1 [23.5, 28.9] & 25.5 \\
Semantic & 55.9 & 3.1 [2.2, 4.4] & 1.7 & 58.7 & 3.1 [2.2, 4.4] & 1.8 & 97.6 & 26.1 [23.5, 28.9] & 25.5 \\
ARC & 57.0 & 3.3 [2.4, 4.6] & 1.9 & 60.0 & 2.6 [1.8, 3.8] & 1.6 & 97.6 & 26.1 [23.5, 28.9] & 25.5 \\
GDSF & 56.8 & 3.9 [2.9, 5.3] & 2.2 & 59.5 & 2.3 [1.5, 3.4] & 1.4 & 97.6 & 24.7 [22.1, 27.5] & 24.1 \\
SISO & 50.7 & 3.1 [2.2, 4.4] & 1.6 & 51.4 & 2.1 [1.4, 3.2] & 1.1 & 97.6 & 27.0 [24.3, 29.8] & 26.4 \\
\bottomrule
\end{tabular}

\end{table*}

\begin{figure}[t]
  \centering
  \includegraphics[width=\columnwidth]{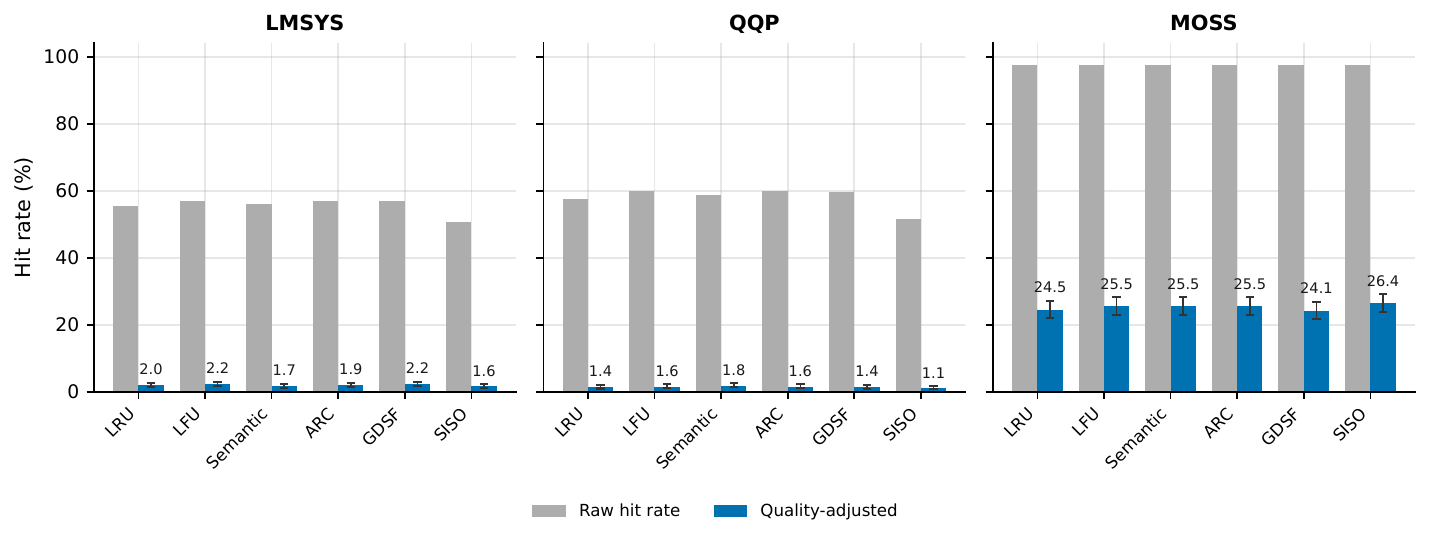}
  \caption{Raw vs.\ quality-adjusted hit rate per dataset and policy.
  Quality adjustment collapses LMSYS and QQP hit rates by more than an
  order of magnitude and erases the ordering among policies.}
  \label{fig:qadj}
\end{figure}

\begin{figure}[t]
  \centering
  \includegraphics[width=\columnwidth]{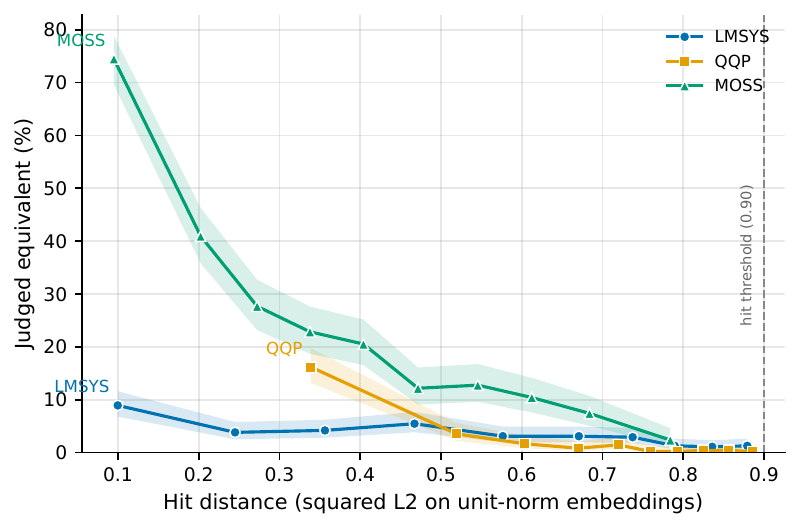}
  \caption{Answer-substitutable rate vs.\ $L_2^2$ hit distance (decile
  bins of unique pairs, Wilson 95\% CIs). MOSS decays steeply; QQP has
  no near-duplicate regime; LMSYS stays low even at near-zero
  distance due to templated prompts.}
  \label{fig:yesdist}
\end{figure}

At the 0.90 hit threshold, Table~\ref{tab:judge} and
Figure~\ref{fig:qadj} show that most matches counted by the
raw hit rate do not answer the incoming query.
On LMSYS, raw hit rates of 50.7--57.0\% shrink to quality-adjusted
rates of 1.6--2.2\% (judged-YES fractions of 3.1--3.9\%); on QQP,
51.4--60.0\% shrinks to 1.1--1.8\% (YES 2.1--3.1\%). Even on MOSS, the
clustered workload where every policy hits 97.6\%, only 24.7--27.0\%
of sampled hits are judged answer-substitutable, for quality-adjusted rates of
24.1--26.4\%.

Quality adjustment also flattens the policy ordering. Within each
dataset, the Wilson 95\% intervals on the YES fractions of all six
policies overlap (Table~\ref{tab:judge}); this sample does not establish a
quality difference among policies. The
inversion on MOSS is illustrative: streaming SISO, the worst policy by raw hit
rate on the other datasets, attains the numerically highest YES rate
(27.0\%). The overlapping intervals do not support a quality ranking.

Figure~\ref{fig:yesdist} shows the mechanism behind these numbers:
answer substitutability declines with embedding distance, and the shape of
the decay differs across datasets. MOSS falls from
74.5\% YES in its nearest distance decile to 2.4\% in its farthest.
QQP has no near-duplicate regime at all: its nearest decile already
sits at median $L_2^2 = 0.34$ with only 16.2\% YES. LMSYS is the
starkest case. Even its nearest decile (median $L_2^2 = 0.10$) yields
only 8.9\% YES, with mid-range deciles at 3--5\%. Manual inspection
attributes this to MiniLM saturating on long templated prompts:
pairs at near-zero distance share an instruction template
(e.g., ``You are the text completion model\ldots'') but differ in the
payload the template wraps, and the judge rejects them. Near-zero embedding
distance therefore does not guarantee answer substitutability.

The verdicts are consistent across two runtimes serving the same judge family.
On the 1{,}931-pair
overlap between the local Ollama judge and Groq's hosted
\texttt{llama-3.1-8b-instant}, raw agreement is 98.55\% with Cohen's
$\kappa = 0.826$; the 28 disagreements concentrate at close distances
(median $L_2^2$ 0.236 vs.\ 0.606 for agreements). Because 1{,}923 of
the overlap pairs are from LMSYS, this consistency check says little about
QQP or MOSS and does not replace human validation.

\section{Discussion: When Admission Suppresses Redundancy}
Insert-on-miss admission places a precise limit on the redundancy signal
available to an eviction policy. The measured density supports that mechanism
in one reference configuration.

\subsection{Packing Condition}
Let $C_t$ be the resident embeddings before query $q_t$, and let $h$ be the
hit radius. With exact lookup and insert-on-miss, $q_t$ is inserted only if
\begin{equation}
\min_{c\in C_t} d(q_t,c)>h.
\end{equation}
For any redundancy radius $r\leq h$, the inserted entry therefore has no
redundancy edge to a resident entry. Eviction cannot create an edge between
unchanged embeddings. This packing condition applies after prefill and does
not cover approximate-search false misses, insertion on hits, or policies that
move stored centroids or otherwise update representations.

\subsection{The Sparsity of the Redundancy Graph}
\begin{figure}[t]
  \centering
  \includegraphics[width=0.88\columnwidth]{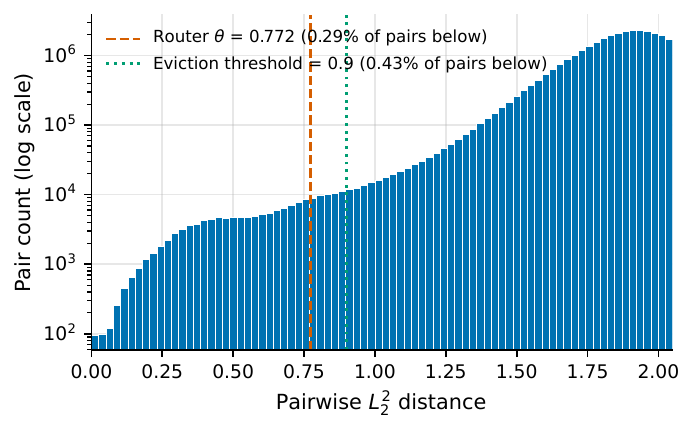}
  \caption{Pairwise $L_2^2$ distance distribution (log-scale
  counts) over an 8{,}572-query subset of the processed LMSYS
  corpus under MiniLM. Pairs within the router or eviction
  threshold are rare, so the redundancy graph over cached
  entries is sparse.}
  \label{fig:distance}
\end{figure}

The processed LMSYS corpus has weak local clustering under the routing and
eviction thresholds used in our experiments. In an 8{,}572-query subset of
the processed corpus, only 0.43\% of the 36.7M pairwise $L_2^2$ distances
fall at or below the 0.90 eviction threshold, and 0.29\% at or below the
router threshold (Fig.~\ref{fig:distance}).

Across all semantic runs (3 seeds $\times$ cache sizes 10\%, 20\%, 30\%),
mean redundancy ranges from $2.2\text{--}2.9\times10^{-5}$, with maximum
observed redundancy 0.035. These values sit below $\mu=0.1$, so the smoothing
term controls the numerator for most entries.

\subsection{Cache-Content Density Characterization}
To locate where the redundancy signal vanishes, we profile density over the
cache contents rather than the workload: at periodic snapshots of the
reference LRU cache during the stream ($10\%$ capacity, MiniLM), we measure
the mean fraction of cached entries within $L_2^2$ distance $\theta$ of each
cached entry.

\begin{figure}[t]
  \centering
  \includegraphics[width=\columnwidth]{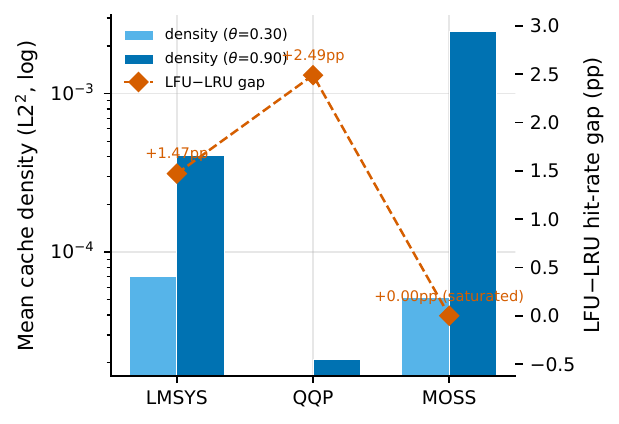}
  \caption{Mean cache-content density (log scale) at $L_2^2$
  $\theta = 0.30$ and $0.90$, with the LFU$-$LRU hit-rate gap
  (right axis) at $10\%$ capacity in the MiniLM LRU profile.}
  \label{fig:density}
\end{figure}

Figure~\ref{fig:density} shows the result. At the eviction threshold
$\theta = 0.90$, cache-content density orders QQP $<$ LMSYS $<$ MOSS
($2.1\times10^{-5}$, $4.1\times10^{-4}$, $2.5\times10^{-3}$ at the final
checkpoint); at $\theta = 0.30$, QQP's cache density falls to zero. QQP was
chosen because its source pairs contain related and duplicate questions, and
it shows the largest
LFU--LRU frequency payoff ($+2.49$pp, versus $+1.47$pp for LMSYS and
$+0.00$pp for saturated MOSS). Yet its cache contents are the
sparsest, $20\times$ below LMSYS and two orders of magnitude below MOSS.

The packing condition predicts low redundancy when the semantic-policy radius
equals the hit radius. Even the largest measured density
($2.5\times10^{-3}$ for MOSS) is $40\times$ below the smoothing constant
$\mu=0.1$, leaving redundancy with little weight in this profile. The profile
covers LRU, 10\% capacity, and MiniLM; it is supporting evidence for the
mechanism, not a measurement of every policy, capacity, or encoder.

\subsection{Mathematical Degradation to LFU}
When $r(e) = 0$, the numerator of our scoring function collapses to the smoothing constant $\mu$. The equation simplifies to:
\begin{equation}
\mathit{score}(e) = \frac{\mu}{\mathit{utility}(e)}
\end{equation}
Because $\mu$ is a constant, the cache entries are ranked entirely by inverse-utility. The semantic policy then behaves exactly like a slightly heavier LFU/LRU hybrid.

\subsection{The Necessity of $\mu$-Smoothing}
\label{sec:mu}
The smoothing term keeps the score defined when redundancy vanishes. Without
$\mu$, an entry with $r(e)=0$ receives score 0 and can remain resident
regardless of its access history. With $\mu>0$, the policy falls back to its
recency-frequency utility.

\subsection{Which Policy Should a Deployment Use?}
\label{sec:guidance}

The measurements support a short decision procedure.

\paragraph{Use LFU as the simple baseline.} LFU is best or within
0.041\,pp of the best policy in every matrix setting and is among the
cheapest. ARC matches it within 0.01\,pp but does not improve on it here.

\paragraph{Treat capacity and policy as separate decisions.} Larger capacities
raise hit rate for every policy and sharply narrow the LFU--LRU gap. Streaming
SISO and FIFO still trail by up to 1.40 and 1.22\,pp at 30\%. We do not
compare the dollar cost of added capacity with engineering cost, so the sweep
does not by itself prescribe provisioning.

\paragraph{Measure resident density before semantic eviction.} Workload-level
similarity does not guarantee that admitted entries remain neighbors. Measure
resident cache density at the intended hit and redundancy radii, and compare
it with the policy's smoothing term. Our MiniLM LRU profile is sparse under
this test; other cache architectures may not be.

\paragraph{Count policy overhead.} The semantic policy costs
$5.83$--$8.24\times$ LRU's per-query time on the full LMSYS runs, and
streaming SISO costs $2.8$--$5.2\times$ across the matrix's sparse cells.
Geometry maintenance is paid even when it does not
change the victim ranking.

\paragraph{Quality-calibrate the hit threshold.} Quantile matching is enough to
avoid a degenerate comparison, not enough to deploy a cache. Select the
threshold on held-out, human-labeled answer-substitutability data and audit
served hits after deployment.

\subsection{What Quality Adjustment Means for the Policy Comparison}
\label{sec:qadj-meaning}
The YES intervals overlap across the six audited policies, so this sample does
not establish a quality difference among them. At the 0.90 threshold, the
judge declines answer reuse for roughly 96--98\% of sampled LMSYS and QQP
hits. The hit threshold and matching model therefore determine how much useful
reuse exists before eviction is considered.
Verified caching reaches the same conclusion from the design side,
replacing the single global threshold with per-entry thresholds under an
explicit error budget \cite{schroeder2025vcache}.
This finding also bears on the router's quality proxy. The
router scores hits by cosine similarity ($1 - d^2/2$), yet the
templated-prompt failure shows that cosine similarity can approach 1.0
on pairs for which the judge rejects answer reuse, so similarity is an unreliable
stand-in for answer quality on boilerplate-heavy text. Consequently, the
60.33\% nominal latency savings (Fig.~\ref{fig:router}) use a cosine
operating point that the judge audit does not directly evaluate, since the audit
fixes the eviction hit threshold ($0.90$ $L_2^2$) rather than the tighter
router-selected $\theta = 0.772$. A distance-conditioned reanalysis of the
existing LMSYS pairs gives context but does not replay cache dynamics at the
router threshold. Restricting the deduplicated judged pairs to distances at
or below the router threshold yields an answer-substitutable rate of 4.5\%
($n = 3{,}912$; Wilson 95\% CI 3.9--5.1\%), against 3.5\% for the full
0.90-threshold sample
(\texttt{scripts/21\_judge\_router\_threshold.py --reanalyze-existing}).
This conditional result suggests that a stricter numeric threshold alone may
not solve the template failure. A response-grounded replay at
$\theta=0.772$ is still required before treating the nominal latency savings
as quality-preserving.

\section{Limitations}

The eviction inputs are ordered, deduplicated corpora rather than production
request traces. They exclude exact repetition, and LMSYS contributes only the
first user utterance from each conversation. The capacity protocol prefills
the first $C$ fraction of the corpus, omits the remainder of the first 30\%, and
measures the final 70\%; it is not a continuous online warmup. The three seeds
vary policy-side randomized sampling, not traces. Our density profile covers
LRU at 10\% capacity under MiniLM, and the streaming-SISO result covers our
adaptation rather than the authors' offline reference implementation.

The matrix uses HNSW for every policy. This controls the backend, but the
harness index configuration's separately measured 2.2\,pp static Recall@1
loss (0.978 at $M{=}32$, ef 128 on 499K vectors) exceeds the differences among
the leading policies, and a false miss changes later cache state. Exact-index
replication is needed for sub-point rankings. Both encoders are compact
English sentence models, and CPU timing may not transfer to GPU serving.
The gte threshold is calibrated on the evaluated embedding distributions, not
held-out quality labels.

The judge audit is directional and response-free. It uses one 8B model family
without human labels; cross-runtime agreement checks consistency only. The
sample covers MiniLM, seed 42, 10\% capacity, and six policies, so its
quality-adjusted rates do not extend to the full matrix. A deployment study
needs stored responses, human answer-substitutability labels, multiple request
traces or time windows, and a held-out threshold-calibration set.

\section{Conclusion}
In the evaluated protocol, no policy improves on LFU by more than
0.041\,pp. Poor choices still matter: FIFO and streaming SISO lose up to
8.67 and 8.55\,pp at 10\% capacity, and streaming SISO costs several times
more per query. LFU is therefore the supported simple baseline, not evidence
that all eviction policies are interchangeable.

The packing condition explains why geometry-aware replacement has little new
signal in an exact insert-on-miss cache when its redundancy radius does not
exceed the hit radius. The density profile is consistent with that mechanism
in one configuration. Exact-index replications and request traces with real
repetition are needed to test how far it extends.

The immediate systems risk lies at admission. Numeric thresholds do not
transfer across encoders, and the evaluated MiniLM operating point admits few
answer-substitutable LMSYS or QQP pairs. The next experiment should combine
held-out human labels, stored responses, exact search, and multiple request
windows. Only then can sub-point policy rankings support a deployment choice
or explain why Sun et al.'s FIFO ordering \cite{sun2026solar} reverses here.
\section{Reproducibility and Artifact Availability}
All code and configuration files are maintained in a version-controlled repository, and the raw result artifacts behind every reported number are retained alongside it with per-run provenance manifests; both are available from the authors on request. Each experiment is produced by a numbered driver script under \texttt{scripts/} paired with a YAML config under \texttt{configs/}, and every run writes a JSON manifest recording its exact command line, the config file and a content hash of its resolved parameters, the Python version (3.11.13) and the package versions active for that run (for the eviction matrix: \texttt{numpy}~2.4.2, \texttt{torch}~2.10.0, \texttt{faiss}~1.14.1, \texttt{transformers}~5.1.0, \texttt{sentence-transformers}~5.2.2), the Git commit, branch, and working-tree state, the node hardware (logical/physical cores, RAM, kernel), and a UTC timestamp.

\textbf{Pipeline.} The core results are regenerated by \path{03_run_index_benchmark.py} (\path{configs/index_benchmark.yaml}); \path{06_run_routing_eval.py} \texttt{--config} \path{configs/routing.yaml} \texttt{--multi-seed} (five-seed adaptive router); \path{08_run_eviction.py} \texttt{--config} \path{configs/eviction.yaml} \texttt{--multi-seed} (three-policy full-LMSYS eviction); the seven-policy matrix via \path{08_run_eviction.py} \texttt{--size 100k} \texttt{--multi-seed}, with \texttt{--dataset} set to each of \texttt{lmsys}, \texttt{qqp}, and \texttt{moss} and \texttt{--embedding-model} to each encoder, against \path{configs/eviction.yaml} (MiniLM) and \path{configs/eviction_gte.yaml} (recalibrated gte-base), the FIFO cells produced by the same driver invoked with \texttt{--policies fifo}; and the judged-quality audit via \path{16_sample_hits_for_judge.py} $\rightarrow$ \path{17_run_llm_judge.py} $\rightarrow$ \path{18_visualize_judge.py}.

\textbf{Claim-to-artifact map.} Table~\ref{tab:artifacts} maps each headline claim to the artifact that produces it.

\begin{table}[h]
  \caption{Claim-to-artifact map. Paths are relative to the repository root.}
  \label{tab:artifacts}
  \footnotesize
  \begin{tabular}{@{}>{\raggedright\arraybackslash}p{0.42\columnwidth}>{\raggedright\arraybackslash}p{0.50\columnwidth}@{}}
    \toprule
    Claim & Artifact \\
    \midrule
    Index Pareto frontier (Fig.~\ref{fig:hnsw}) & \path{results/benchmarks/index_benchmark_*.json} \\
    Adaptive routing, 60.33\% savings at $\theta{=}0.772$ (Fig.~\ref{fig:router}) & \path{results/routing/routing_eval_multi_seed.json} \\
    Three-policy full-LMSYS eviction (Fig.~\ref{fig:eviction}, Table~\ref{tab:hit-rate}) & \path{results/eviction/eviction_results_multi_seed.json} \\
    Seven-policy matrix (Figs.~\ref{fig:heatmap},~\ref{fig:ablation}) & \path{results/eviction/phase6_matrix_*}, \path{results/eviction/phase10_fifo_*} \\
    Recalibrated gte-base replication (Sec.~\ref{sec:calibration}) & \path{results/eviction/phase6_recal_*} \\
    Judged hit quality (Table~\ref{tab:judge}, Figs.~\ref{fig:qadj},~\ref{fig:yesdist}) & \path{results/judge/}, \path{paper/tables/phase8_judge_table.tex} \\
    Router-threshold judged quality (Sec.~\ref{sec:qadj-meaning}) & \path{results/judge/phase7_minilm_c0p10_local/verdicts.jsonl} at $L_2^2 \le 0.772$ \\
    \bottomrule
  \end{tabular}
\end{table}

\textbf{Repeatability and provenance.} Eviction cells run seeds 42, 123, and
456, and routing uses 42, 123, 456, 789, and 1024. The maximum eviction
hit-rate standard deviation is $3.1\times10^{-4}$ ($2.2\times10^{-4}$ for
MiniLM); these seeds vary internal randomized sampling rather than request
traces. The capacity-sized prefill, fixed final-70\% suffix, and per-encoder
thresholds are recorded in configuration and manifests. Representative run
commits are \texttt{39676dd} (routing), \texttt{e4f39ee} (three-policy
eviction), and \texttt{d386ffa} (policy matrix); FIFO cells record their
commits in their manifests. The index benchmark predates manifest
instrumentation and is regenerated from
\texttt{03\_run\_index\_benchmark.py} and its configuration.

\section{Ethical Considerations}
This work uses the publicly released LMSYS-Chat-1M, Quora Question Pairs, and MOSS datasets and does not involve new human-subject data collection. We evaluate systems-level cache behavior rather than user-level profiling, and report aggregate metrics only. The LLM judge runs locally on open-weight models; no query content was sent to third-party services beyond a rate-limited validation subset.

\section*{Acknowledgments}
This project was conducted as part of CSE 584 (Advanced Database Systems) at the University of Michigan. We thank Professor Lin Ma for his guidance and feedback throughout the course and this project. We also thank the creators of the LMSYS-Chat-1M, Quora Question Pairs, and MOSS datasets, and acknowledge the University of Michigan Great Lakes HPC cluster, which ran the large-scale eviction experiments.
\vspace{-0.3cm}
\balance
\bibliographystyle{plain}
\bibliography{references}
\end{document}